\documentclass[11pt,aps,pra,amsmath,amssymb]{revtex4-1}
\usepackage{graphicx}
\usepackage{siunitx}
\usepackage{amsmath}
\usepackage{color}
\usepackage{booktabs}
\AtBeginDocument{
	\heavyrulewidth=.08em
	\lightrulewidth=.05em
	\cmidrulewidth=.03em
	\belowrulesep=.65ex
	\belowbottomsep=0pt
	\aboverulesep=.4ex
	\abovetopsep=0pt
	\cmidrulesep=\doublerulesep
	\cmidrulekern=.5em
	\defaultaddspace=.5em
}

\def\bea{\begin{eqnarray}}
\def\eea{\end{eqnarray}}
\def\ben{\begin{equation}}
\def\een{\end{equation}}
\def\benu{\begin{enumerate}}
\def\enu{\end{enumerate}}
\def\beal{\begin{equation}\begin{aligned}} 
\def\eeal{\end{aligned}\end{equation}}

\def\sss{\scriptscriptstyle\rm}

\def\hatT{{\hat T}}
\def\hatV{{\hat V}}
\def\hatH{{\hat H}}

\def\br{{\mathbf r}}
\def\brp{{\mathbf r}^\prime}

\def\bk{{\mathbf k}}
\def\bq{{\mathbf q}}

\def\bj{{\mathbf j}}
\def\bF{{\mathbf F}}

\def\diff{{\rm d}}
\def\d3{{\rm d^3}}

\def\intdbr{\int\diff{\br}}
\def\intdbrs{\intdbr\,}

\def\s{_{\sss S}}
\def\b{_{\sss B}}
\def\p{_{\sss p}}
\def\P{_{\sss P}}

\def\xc{_{\sss XC}}

\def\Hxc{_{\sss HXC}}

\def\H{_{\sss H}}

\def\VW{^{\rm vW}}
\def\vw{{von Weizs\"acker}}

\def\kin{^{\rm kin}}
\def\intss{^{\rm int}}

\usepackage{xspace}

\usepackage{mathtools}

\newcommand{\braket}[2]{\left\langle #1 \middle\vert #2 \right\rangle}
\newcommand{\brakett}[3]{\left\langle #1 \middle\vert #2 \middle\vert #3 \right\rangle}

\newcommand{\eqn}[1]{\mbox{Eq.\hspace{1pt}(\ref{#1})}}

\begin{document}

\title{Time-dependent Orbital-free Density Functional Theory: Background and Pauli kernel approximations}

\author{Kaili Jiang}
\email{kaili.jiang@rutgers.edu}
\affiliation{Department of Chemistry, 73 Warren St., Rutgers University, Newark, NJ 07102, USA}

\author{Michele Pavanello}
\email{m.pavanello@rutgers.edu}
\affiliation{Department of Chemistry, 73 Warren St., Rutgers University, Newark, NJ 07102, USA}
\affiliation{Department of Physics, 101 Warren St., Rutgers University, Newark, NJ 07102, USA}

\begin{abstract}
	Time-dependent orbital-free DFT is an efficient method for calculating the dynamic properties of large scale quantum systems due to the low computational cost compared to standard time-dependent DFT. We formalize this method by mapping the real system of interacting fermions onto a fictitious system of non-interacting bosons. The dynamic Pauli potential and associated kernel emerge as key ingredients of time-tependent orbital-free DFT. Using the uniform electron gas as a model system, we derive an approximate frequency-dependent Pauli kernel. Pilot calculations suggest that space nonlocality is a key feature for this kernel. Nonlocal terms arise already in the second order expansion with respect to unitless frequency and reciprocal space variable ($\frac{\omega}{q\, k_F}$ and $\frac{q}{2\, k_F}$, respectively). Given the encouraging performance of the proposed kernel, we expect it will lead to more accurate orbital-free DFT simulations of nanoscale systems out of equilibrium. Additionally, the proposed path to formulate nonadiabatic Pauli kernels presents several avenues for further improvements which can be exploited in future works to improve the results.
\end{abstract}

\date{\today}
\maketitle

\section{Introduction}
The simulation of large scale quantum systems (such as nanometer-scale quantum dots) when their electrons are out of equilibrium has been a challenging undertaking and cause for frustration \cite{SD07,HCM08,MLB08,SD08,Vignale08}. The challenge is multifaceted. First, the simulation methods need to be predictive, and although time-dependent DFT (TD-DFT) is the workhorse for these types of simulations, it still lacks broad applicability across the possible excited states characters (valence, charge transfer, Rydberg)
\cite{TAH+99,PBH+08,DWH03,CH08a,CH08,LB94,RMP16,WKV09}. Second, the computational cost is a major concern. The cubic scaling of the ground state DFT algorithm and also of the real-time or linear-response TD-DFT algorithms (provided a small number of states are computed) cripples the applicability of these methods to nanoscale systems \cite{YNN+18,LH16}. The problem is further exacerbated when systems with dense spectra are considered \cite{AA18, Neugebauer10, Pavanello13}. Beyond DFT, an array of accurate methods is available. However, their computational cost is typically orders of magnitude larger \cite{GLL18}. 

In this work, we tackle issues of computational feasibility by exploring an alternative path: orbital-free DFT (OF-DFT) \cite{WC02, WW13}. OF-DFT effectively reduces the complexity of the problem by considering only a single active orbital. This is in stark contrast with commonly adopted Kohn-Sham DFT (KS-DFT) methods in which a set of occupied orbitals equal in number to the electrons in the system needs to be considered \cite{HK64,KS65}. Thus, OF-DFT massively reduces the complexity of the problem, provided that accurate approximations for the density functionals involved are available. An important feature distinguishing OF-DFT from KS-DFT is that in addition to the exchange-correlation (XC) functional, OF-DFT also requires the knowledge of the noninteracting kinetic energy functional.

Employing currently available density functional approximants, the ground state version of OF-DFT scales linearly with the system size and has been shown to be applicable to main group metals and III-V semi-conductors \cite{WGC99,HC10,SC14a,CFD17,LKT18}. Recently, non-local functionals such as HC \cite{HC10} and LMGP \cite{MP19} have achieved chemical accuracy for a wider range of systems.

OF-DFT can be formulated in the time domain to approach systems out of equilibrium \cite{GVV+76,PBN+86,DRS98,RLB00,ZT94,ZZ99,BH00,MGB15,MBR18,Yan15,NPC+11,ZXZ+17,DWH+18}. It is sometimes referred to as the time-dependent Thomas-Fermi method \cite{PBN+86,DRS98,RLB00} and also hydrodynamic DFT \cite{BH00,MGB15,MBR18,Yan15}. In this work, we will refer to it as TD-OF-DFT. Practical implementations initially featured the adiabatic Thomas-Fermi (TF) \cite{Thomas26,Fermi27}  plus \vw\ (vW) \cite{Weizsaecker35} approximation \cite{GVV+76,DRS98} (TFW, hereafter), and later including nonadiabatic corrections in the potential \cite{NPC+11,ZXZ+17,DWH+18}. TD-OF-DFT has seen a wide range of applications including atoms and clusters in laser fields, electron dynamics and optical response in nano-structures, and oscillators in electric fields \cite{ZGS88,PSM93,DRS98,BH08,PB15,RLB00,CKC15}. Despite the drastic approximations, TD-OF-DFT has been quite successful in describing the optical spectra of metal clusters \cite{DRS98,SJM+20}.

In this work we propose an alternative formalism of TD-OF-DFT by introducing a map between the real system and a fictitious system of noninteracting bosons. We show that this formalism is in principle exact by following proofs to theorems similar to the ones at the foundation of regular TD-DFT. The resulting boson-like singe-orbital representation is a compact and computationally amenable representation of the density and velocity field
in a single complex object. Thus, TD-OF-DFT accomplishes a similar goal as the Madelung wavefunction typical of hydrodynamic DFT \cite{BH00,MGB15,MBR18,Yan15} and other treatments \cite{Madelung_1927}. However, the Madelung wavefunction is ad-hoc for multi-electron systems and its use involves making the approximation that the electronic structure of a many-electron system can be described by a function of a single, collective variable. Such approximations are not invoked in TD-OF-DFT which is an exact formalism.

However exact, TD-OF-DFT's formalism does not provide practically useable expressions for the time-dependent potentials involved. Approximations to these potentials (and specifically the Pauli potential contribution to the total time-dependent potential) are the subject of this work and finding good approximations is key to the accuracy of the method. We should stress that our undertaking is not an academic exercise, it is rather a search for time-dependent electronic structure methods that are much cheaper computationally to currently available methods. In this work we formulate exact conditions which are a foundational aspect of functional development \cite{SRP15} and focus particularly on the TD-OF-DFT Pauli kernel, its nonadiabatic behavior, spatial nonlocality and approximations needed for practical calculations.

To cast our developments in the current state of the art, we should mention other quantum dynamic methods, such as time-dependent density functional tight binding (TD-DFTB) \cite{PFK+95,TSZ+11,ARJ+20,ASS+19}, as well as simplified versions of TD-DFT \cite{WSG20,WSC+19,WG19,WG18} which are capable of handling systems of much greater size than conventional TD-DFT. 

This work is organized as follows. In section \ref{sec:boson}, we first establish an exact map between the real system of interacting electrons, a system of noninteracting electrons and a system of noninteracting bosons. This allows us to formulate TD-OF-DFT as an exact formalism with theorems and exact properties/conditions for the building blocks of the method. In section \ref{sec:rt} we derive the real-time formalism of this approach (i.e., the Schr\"{o}dinger-like equation), and the Pauli potential that needs to be approximated in actual calculations, as well as some properties of the exact Pauli potential. In section \ref{sec:lr} we focus on the linear response formalism and derive Dyson equations relating the response to external perturbations of the real system, the KS system and the fictitious noninteracting boson system.
We then proceed to uncover a route to approximate the Pauli kernel recovering needed nonadiabaticity and nonlocality. Pilot calculations are also presented to showcase the newly developed Pauli kernel.

\section{Noninteracting Boson system}\label{sec:boson}
The foundation of ground state Kohn-Sham (KS)-DFT and TD-DFT is the existence of a unique and invertible map between the real system of interacting fermions and a fictitious system of noninteracting fermions (KS system, hereafter). However, other unique and invertible maps can be found between the real system and other fictitious systems. For example, it is possible to use a system of bosons having the same charge, point-like shape and mass as the electrons. Such a bosonic system is fictitious, and thus does not need to be rooted in reality. A bosonic system at zero temperature yields the following wavefunction to density relationship,
\begin{equation}
\label{eq:boson_den}
n(\br)=N|\phi\b(\br)|^2.
\end{equation}

OF-DFT implicitly takes advantage of the fictitious bosonic system because it can be formulated in a way that reduces to \eqn{eq:boson_den}, i.e., only utilizes a single active orbital. This can be seen by invoking the KS-DFT total energy functional
\begin{equation}
	\label{}
	E[n]=T\s[n] + E\H[n] + E\xc[n] + \intdbrs n(\br)v(\br),
\end{equation}
where $T\s$ is the kinetic energy of $N$ noninteracting electrons having a density of $n(\br)$. $E\H$ is the Hartree energy, $E\xc$ is the XC energy, and $v(\br)$ is the external potential.
A search over electron densities to find the minimum of the energy functional must be carried out with the constraint that the densities must integrate to $N$ electrons. Thus, the following Lagrangian is typically invoked,
\begin{equation}
	\label{eq:lag_ofdft}
	\mathcal{L}[n] = E[n] - \mu \bigg[\intdbrs n(\br) - N\bigg].
\end{equation}
Formally, the minimum of the above Lagrangian is also a stationary point \cite{Levy79} and therefore its functional derivative with respect to the electron density must vanish,
\begin{equation}
	\label{}
	0=\frac{\delta T\s[n]}{\delta n(\br)} + v\s(\br) - \mu,
\end{equation}
which is known as the Euler equation of DFT. If the first term on the RHS of the above equation is written as the vW potential, $v\VW(\br)=\frac{1}{\sqrt{n(\br)}}\left(-\frac{1}{2}\nabla^2 \sqrt{n(\br)}\right)$, plus a correction, $\frac{\delta T\s[n]}{\delta n(\br)}=v\VW(\br) + \left( \frac{\delta T\s[n]}{\delta n(\br)} - v\VW(\br)\right)$, then because the vW potential is exact for wavefunctions of up to two electrons and for bosonic systems, the correction term is known as Pauli potential, $v\P(\br)$, as it accounts for Fermi statistics. 

The single-orbital wavefunction now emerges as simply $\phi\b(\br)=\frac{1}{\sqrt{N}}\sqrt{n(\br)}$ and the following Schr\"odinger-like equation derives directly from \eqn{eq:lag_ofdft},
\begin{equation}
	\label{eq:schrodinger_ofdft}
	\bigg( -\frac{1}{2}\nabla^2 + \underbrace{v\P(\br) + v\s(\br)}_{v\b(\br)}\bigg) \phi\b(\br) = \mu \phi\b(\br).
\end{equation}

Following \eqn{eq:boson_den}, the many-body wavefunction of the noninteracting boson system is a Hartree product:
\begin{equation}\label{eq:boson_wf}
\Phi\b(\br_1,\br_2,\dots,\br_N,t)=\prod_l^N\phi\b(\br_l,t).
\end{equation}
In this article we use the subscript $\mathrm{B}$ for the noninteracting boson system, subscript $\mathrm{S}$ for the KS system, and use no subscript for the real system.

\begin{table}[h]
	\caption{Comparison between the interacting, KS and noninteracting boson system.}
	\label{tab:3sys}
	\centering
	\begin{tabular}{l l l l}
		\toprule
		& \begin{tabular}{@{}l@{}} Interacting\\ system \end{tabular} & KS system & \begin{tabular}{@{}l@{}} Noninteracting\\ boson system \end{tabular}\\\midrule
		density & $n(\br,t)$ & $n(\br,t)$ & $n(\br,t)$\\\midrule
		wavefunction & $\Psi(\br_1,\dots,\br_N,t)$ & $\frac{1}{\sqrt{N!}}\det[\{\phi\s{}_{,l}(\br_l,t)\}]$ & $\prod_l^N\phi\b(\br_l,t)$ \\\midrule
		\begin{tabular}{@{}l@{}} effective\\ potential \end{tabular} & $v(\br,t)$ &$\begin{array}{l} v\s[n](\br,t)=\\v(\br,t)+v\H[n](\br,t)\\+v\xc[n](\br,t) \end{array}$ & $\begin{array}{l}v\b[n](\br,t)=\\v\s[n](\br,t)+v\P[n](\br,t) \end{array}$ \\\midrule
		Hamiltonian & $\hatT + \hat{W} +v(\br,t)$ & $\hatT +v\s(\br,t)$ & $\hatT +v\b(\br,t)$ \\\bottomrule
	\end{tabular}
\end{table}

When considering systems away from equilibrium (i.e., when the density, wavefunction and potentials become time-dependent) the DFT theorems remain largely valid \cite{Ullrich11}. Before going in to the details of the formal proofs, we summarize in Table \ref{tab:3sys} the most important quantities involved in the three reference systems (interacting, KS and noninteracting boson). The wavefunctions of the three systems are different. For the interacting system it can be a very complex function of coordinates of every electron and time; for the KS system it is a single Slater determinant; and, as mentioned, for the noninteracting boson system it is a simple Hartree product of the same function, $\phi_B$. The interacting system features an electron-electron Coulomb repulsion term $\hat{W}$ in the Hamiltonian, whereas the two noninteracting systems do not. To compensate for the missing electron-electron interaction, the KS system employs an effective potential, $v\s[n](\br,t)$, which includes the external potential $v(\br,t)$, the Hartree potential, $v\H[n](\br,t)$, and the XC potential, $v\xc[n](\br,t)$. For the noninteracting boson system, there is a different effective potential, $v\b[n](\br,t)$, which includes an additional Pauli potential term, $v\P[n](\br,t)$, required to compensate for the neglect of the Pauli exclusion principle. In the end, all three systems yield the same time-dependent density. 

\begin{figure}[htp]
	\centering
	\includegraphics[width=0.6\textwidth]{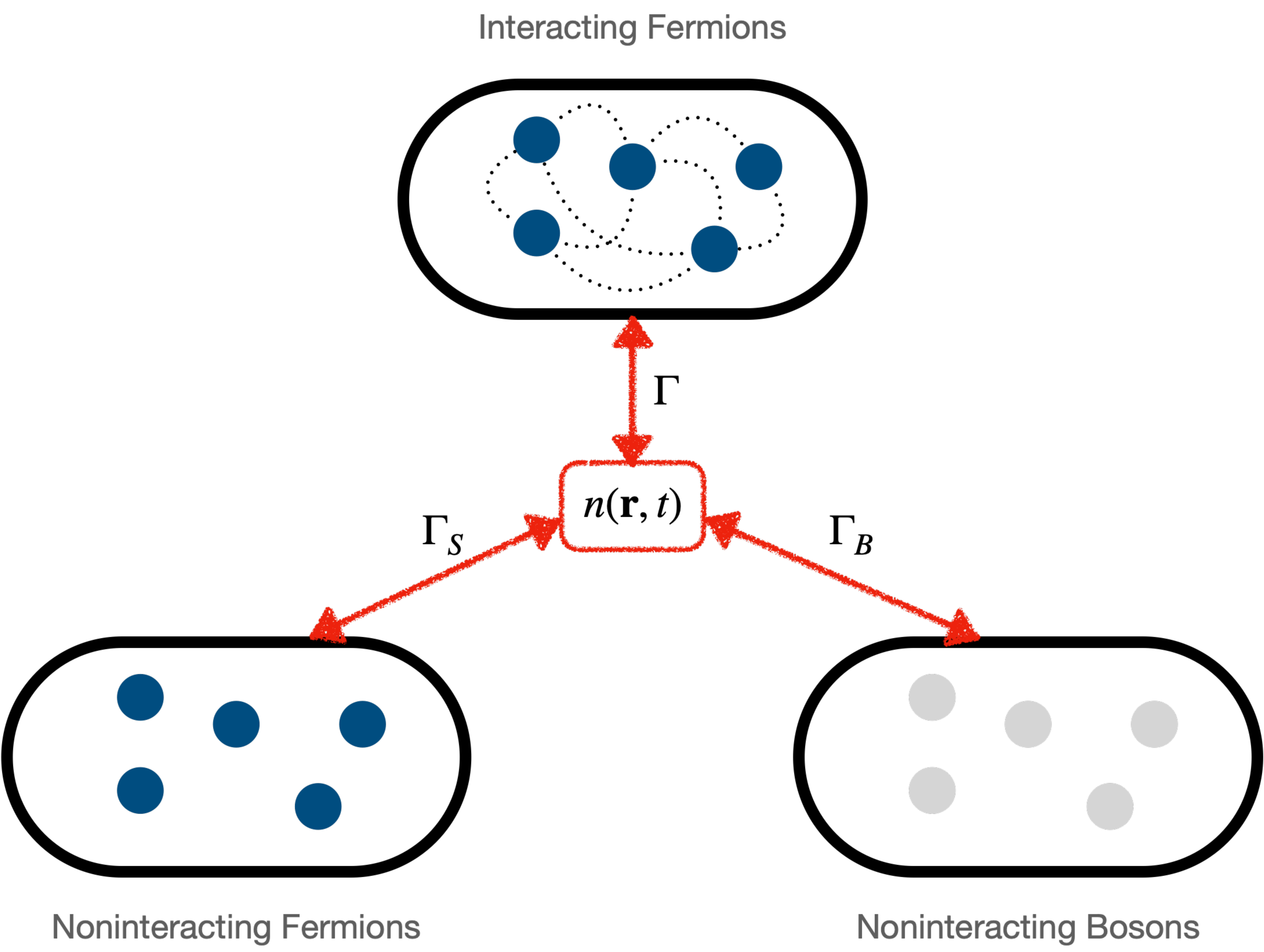}
	\caption{Three representations of an electronic system discussed in this work. The circles represent electrons and the dotted lines represent electron-electron interactions. The noninteracting fermion and boson systems are depicted in the bottom left and right enclosures, respectively, while the interacting (real) system is at the top. Grey circles represent bosons. All three systems have a one to one bijective map between its wavefunction, effective potential and time-dependent density, $n(\br,t)$ which are indicated by the symbols $\Gamma$, $\Gamma\s$ and $\Gamma\b$. The density is the same across all representations.}
	\label{fig:mapping}
\end{figure}

To provide a visual and formal framework, we introduce bijective maps connecting the three systems and realized in practice by the $v\s$ and $v\P$ potentials. We shall indicate with $\Gamma$, $\Gamma\s$ and $\Gamma\b$ bijective maps linking every $v$-representable time-dependent density to a unique time-dependent multi-electron wavefunction $\Psi(\br_1,\dots,\br_N,t)$, every non-interacting $v\s$-representable densities to a unique set of KS orbitals $\{\phi_l(\br_l,t)\}$, every non-interacting $v\b$-representable densities to a unique single orbital $\phi\b(\br,t)$, respectively (see Figure \ref{fig:mapping}). 

To establish the exactness of these maps, we now prove analogs of two fundamental theorems in TD-DFT. The first theorem establishes one-to-one bijective maps between time-dependent densities, time-dependent wavefunctions and time-dependent effective potentials for any one of the systems, particularly for the noninteracting boson system. From \eqn{eq:boson_den} and \eqn{eq:boson_wf}, given a wavefunction the density is uniquely determined, and with a given density, the wavefunction is uniquely determined up to a phase factor. The forward one-to-one map between the time-dependent effective potentials and time-dependent densities is also found by pluging the potential in the corresponding Hamiltonian and solving for the time-dependent Schr\"{o}dinger-like equation. The reverse one-to-one map between the time-dependent effective potentials and time-dependent densities is proved by the Runge-Gross theorem \cite{RG84}. The proof of this theorem does not require the Hamiltonian and the wavefunction of the system to be in a particular form, except for the effective potential to be Taylor expandable \cite{YMB12}. Therefore, it applies to the noninteracting boson system without any effort. 

Next, we are going to establish a one-to-one map between the external potential of the interacting system and the effective potential of the noninteracting boson system (up to a constant). This requires to prove an analog of the van Leeuwen theorem \cite{Leeuwen99}: For any time-dependent density, $n(\br,t)$, associated with the system of interacting fermions and the external potential, $v(\br,t)$, and initial interacting state, $\Psi_0$, there exist a unique potential $v\b(\br,t)$ up to a purely time-dependent constant, $c(t)$, that will reproduce the same time-dependent density of a system of noninteracting bosons, where the initial state $\Phi\b^0$ of this system must be chosen such that the densities and their time derivative of the two systems are the same at the initial time. 

The proof of our analog of the van Leeuwen theorem is very similar to that of the original one. Here we are going to follow Ref.\ \citenum{Ullrich11} and only point out the differences. We start with the Hamiltonian of the two systems, the interacting fermions,
\begin{equation}
	\label{eq:ham_int}
	\hatH = \hatT + \hat{W} +\hatV(t),
\end{equation}
with the time-dependent wavefunction $\Psi(t)$ and initial state $\Psi_0$, and the noninteracting bosons,
\begin{equation}
	\label{eq:ham_boson}
	\hatH\b = \hatT + \hatV\b(t),
\end{equation}
with the time-dependent wavefunction $\Phi\b(t)$ and initial state $\Phi\b^0$. Both systems yields the same time-dependent density $n(\br,t)$. 
We can Taylor expand $v(\br,t)$
\begin{equation}
\label{}
v(\br,t)=\sum_{k=0}^{\infty}\frac{1}{k!}v_k(\br)(t-t_0)^k,
\end{equation}
where $v_k(\br)=\partial^k v(\br,t)/\partial t^k\vert_{t=t_0}$, and $v\b(\br,t)$
\begin{equation}
\label{}
v\b(\br,t)=\sum_{k=0}^{\infty}\frac{1}{k!}v\b^k(\br)(t-t_0)^k,
\end{equation}
where $v\b^k(\br)=\partial^k v\b(\br,t)/\partial t^k\vert_{t=t_0}$.
Our goal is to find a way to uniquely determine each Taylor expansion coefficients $v\b^k(\br)$.

Next, we note that the current density operator should be the same for both systems. Namely, its expectation value
\begin{equation}
	\label{eq:operator_j}
	\hat{\bj}(\br)=\frac{1}{2i}\sum_l^N[\nabla_l\delta(\br-\br_l)+\delta(\br-\br_l)\nabla_l].
\end{equation}
should be the same whether it is evaluated with the KS wavefunction or the bosonic wavefunction.

Thus, applying \eqn{eq:operator_j} to the wavefunctions of the interacting, KS, and noninteracting boson systems we obtain the current densities for these systems, 
\begin{align}
	\label{}
	\bj(\br,t)&=\brakett{\Psi(t)}{\hat{\bj}(\br)}{\Psi(t)},\\
	\label{}
	\bj\s(\br,t)&=\frac{1}{2i}\sum_{l}^{N}[\phi\s{}_{,l}^*(\br,t)\nabla\phi\s{}_{,l}(\br,t)+\phi\s{}_{,l}(\br,t)\nabla\phi\s{}_{,l}^*(\br,t)],\\
	\label{}
	\bj\b(\br,t)&=\frac{N}{2i}[\phi\b^*(\br,t)\nabla\phi\b(\br,t)+\phi\b(\br,t)\nabla\phi\b^*(\br,t)].
\end{align}

If we apply the equation of motion for $\hat{\bj}(\br)$ in the noninteracting boson system, we will get (See Appendix \ref{sec:der_Fkin} for detailed derivation)
\begin{equation}
	\label{eq:motion_j}
	\frac{\partial}{\partial t}\bj\b(\br,t)=-n(\br,t)\nabla v\b(\br,t)-\bF\kin(\br,t).
\end{equation}
Where the kinetic force $\bF\kin$ is in the following form
\begin{equation}
	\label{eq:Fkin}
	\bF\kin(\br,t)=\frac{N}{4}\{4[\nabla\phi\b^*(\br,t)][\nabla^2\phi\b(\br,t)]+4[\nabla^2\phi\b^*(\br,t)][\nabla\phi\b(\br,t)]-\nabla^3[\phi\b^*(\br,t)\phi\b(\br,t)]\}.
\end{equation}
\eqn{eq:motion_j} is in the same form as Eq.\hspace{1pt}(3.27) in Ref.\ \citenum{Ullrich11} if we consider the interaction force $\bF\intss=0$ because the system is assumed to be noninteracting. 
Take the divergence of \eqn{eq:motion_j} and use the continuity equation, we obtain
\begin{equation}
\label{eq:2nd_der_n}
	\frac{\partial^2}{\partial t^2}n(\br,t)=\nabla\cdot[n(\br,t)\nabla v\b(\br,t)]+q\b(\br,t),
\end{equation} 
where $q\b(\br,t)=\nabla\cdot\bF\kin(\br,t)$.
Subtract \eqn{eq:2nd_der_n} from its counter part for the interacting system Eq.\hspace{1pt}(3.48) in Ref.\ \citenum{Ullrich11}, we obtain
\begin{equation}
\label{eq:van_leeuwen_key}
\nabla\cdot[n(\br,t)\nabla\gamma(\br,t)]=\zeta(\br,t),
\end{equation}
where $\gamma(\br,t)=v(\br,t)-v\b(\br,t)$ and $\zeta(\br,t)=q\b(\br,t)-q(\br,t)$.
Note that \eqn{eq:van_leeuwen_key} is in the exactly same form as Eq.\hspace{1pt}(3.50) in Ref.\ \citenum{Ullrich11}.

The last part of the proof is to derive the equations that uniquely determine $v\b^k(\br)$ from \eqn{eq:van_leeuwen_key}. This part of the proof is exactly the same as those from Eq.\hspace{1pt}(3.51) to Eq.\hspace{1pt}(3.55) in Section 3.3 of Ref.\ \citenum{Ullrich11}. 
Now we reach the conclusion which is that $v\b(\br,t)$ is uniquely determined by the density and the initial states, e.g.,
\begin{equation}
	\label{}
	v\b \equiv v\b[n,\Psi_0,\Phi\b^0](\br,t).
\end{equation}


We formally proved that for every interacting system with a $v$-representable time-dependent density and given an initial condition, there is a unique, noninteracting boson system that yields the same density. This is important because without such a map a reference noninteracting boson system cannot be employed. We recall that this approach has been analyzed before by several authors in several regimes under different sets of approximations \cite{SD07,HCM08,DRS98,NPC+11,BH00,DWH+18,WCD+18,ZT94,Schaich93,Yan15,DG82,HD81} as well as in the context of the exact factorization method of recent formulation \cite{SG17}. In the next section, we discuss how to determine the effective bosonic potential in practical calculations and we also review some of its exact properties that can be useful for guiding the development of approximations.

\section{Additional theorems and properties}\label{sec:rt}
\subsection{Time-dependent Schr\"{o}dinger-like equation}

The typical setup is that the system starts in its ground state at $t=t_0$ and begins to evolve under the influence of a time-dependent external potential for $t>t_0$. Thus, the initial wave function, $\Phi\b^0$, can be found with ground-state OF-DFT by solving \eqn{eq:schrodinger_ofdft}. In the following, it will be convenient to define the time-independent Pauli potential as a functional derivative of the Pauli kinetic energy. Namely,
\begin{equation}
v\P[n](\br) = \frac{\delta{T\P[n]}}{\delta n(\br)},
\end{equation}
where $T\P[n]=T\s[n]-T\s\VW[n]$. 

Given the initial orbital $\phi\b^0$,
\begin{equation}
	\label{}
	\phi\b^0(\br)=\frac{1}{\sqrt{N}}\sqrt{n_0(\br)},
\end{equation}
where $n_0(\br)$ is the ground-state density at $t=t_0$, in order to propagate the system after $t>t_0$, we need to solve a time-dependent Schr\"{o}dinger-like equation:
\begin{equation}
\left[-\frac{\nabla^2}{2}+v\b[n](\br, t)\right]\phi\b({\br, t})=i\frac{\partial}{\partial t}\phi\b({\br, t}),
\end{equation}
with the initial condition
\begin{equation}
	\label{}
	\phi\b(\br,t_0)=\phi\b^0(\br)
\end{equation}
and the time-dependent effective potential is given by
\begin{equation}\label{eq:veff_td}
	v\b[n](\br, t) = v\P[n](\br,t) + v\s[n](\br,t),
\end{equation}
which defines the time-dependent Pauli potential. 
   
\subsection{Properties of $v\P(\br,t)$}

While the exact form of the time-dependent Pauli potential is unknown for the general case, we can still derive some properties and exact conditions. \eqn{eq:veff_td} indicates that the role of the time-dependent Pauli potential in TD-OF-DFT is similar to the role of time-dependent XC potential in TD-DFT because it ensures that the fictitious boson system has the same electronic dynamics as the fermion system. In TD-DFT, the XC potential ensures that the dynamics of the noninteracting fermion system is the same as the interacting fermion system. Therefore, for many properties of the time-dependent XC potential, we can find analogies for the time-dependent Pauli potential. Here we are going to list some of such properties.

\subsubsection{Functional dependency}
The functional dependencies of $v\P$, $v\s$ and $v\b$, in terms of $n_0$, $\Psi_0$, $\Phi\s^0$ and $\Phi\b^0$
\begin{align}
	\label{eq:func_s}
	v\s &\equiv v\s[n_0,\Psi_0,\Phi\s^0](\br), \\
	\label{eq:func_b}
	v\b &\equiv v\b[n_0,\Psi_0,\Phi\b^0](\br), \\
	\label{eq:func_p}
	v\P &\equiv v\P[n_0,\Phi\s^0,\Phi\b^0](\br).
\end{align}
We stress that the formal dependency of the time-dependent boson potential in \eqn{eq:func_b} is a direct result of the initial conditions of the time-dependent problem as it arises in the proof of the van Leeuwen theorem in Section \ref{sec:boson}.
\eqn{eq:func_p} links the KS system with the noninteracting boson system. Thus, it does not depend on the interacting wavefunction.

%

\subsubsection{The zero-force theorem}

The total momentum of a many-body system can be expressed as \cite{Ullrich11}
\begin{equation}
	\label{eq:momentum}
	\mathbf{P}(t)=\intdbrs\bj(\br,t).
\end{equation}
Combine \eqn{eq:motion_j} and \eqn{eq:momentum} and notice that $\bF\kin$ in the form of \eqn{eq:Fkin} is the divergence of a stress tensor and thus the integral of $\bF\kin$ over the space vanishes, we obtain
\begin{equation}
	\label{eq:momentum_boson}
	\frac{\partial\mathbf{P}\b(t)}{\partial t}=-\intdbrs n(\br,t)\nabla v\b(\br,t).
\end{equation}

The total momentum of the noninteracting boson system and the KS system are the same at all time because the densities of the two systems are the same, thus
\begin{equation}
	\label{}
	0=\frac{\partial\mathbf{P}\s(t)}{\partial t}-\frac{\partial\mathbf{P}\b(t)}{\partial t}
	=-\intdbrs n(\br,t)\nabla[v\s(\br,t)-v\b(\br,t)]
\end{equation}
or
\begin{equation}\label{}
	\intdbrs n(\br,t)\nabla v\P(\br,t)=0
\end{equation}

The implication of this theorem are that $v\P(\br,t)$ cannot exert a total, net force on the electronic system. This is also the case for the XC potential \cite{Vignale95}.

\subsubsection{The one- or two-electron limit}

Similar to the XC potentials in KS-DFT and TD-DFT, an exact condition for the Pauli potential arises for one- and two-electron densities. In the event of the system having only one electron or two spin-compensated electrons, the KS system is given by a single orbital. This orbital is the same as the orbital of the noninteracting boson system. Thus,
\begin{equation}
	\label{}
	v\P(\br,t)=0 ~~\text{ for } \intdbrs n(\br) \leq 2.
\end{equation}
Even though this is a clear exact condition, its imposition in real-life density functional approximations is nearly impossible resembling the self-interaction error for the XC functional \cite{PZ81}.

\subsubsection{The relation between $T\P$ and $v\P$}

Let us define the time-dependent energy of the noninteracting boson as
\begin{equation}
	\label{eq:td_energy}
	E(t)=\brakett{\Phi\b(t)}{\hat{H\b}(t)}{\vphantom{\hat{H\b}}\Phi\b(t)}.
\end{equation}
We then plug \eqn{eq:ham_boson} and \eqn{eq:veff_td} in \eqn{eq:td_energy} to obtain
\begin{equation}
	\label{eq:e_boson}
	E(t)=T\b(t)+T\P(t)+E\H(t)+E\xc(t)+\intdbrs v(\br,t)n(\br,t),
\end{equation}
where
\begin{align}
	\label{}
	T\b(t)&=-\sqrt{n(\br,t)}\frac{\nabla^2}{2}\sqrt{n(\br,t)},\\
	\label{}
	T\P(t)&=T\s(t)-T\b(t),\\
	\label{}
	E\H(t)&=\frac{1}{2}\intdbr\intdbr'\,\frac{n(\br,t)n(\br',t)}{|\br-\br'|},\\
	\label{}
	E\xc(t)&=T(t)-T\s(t)+W(t)-E\H(t).
\end{align}
Note that the definition of $E\H(t)$ and $E\xc(t)$ are exactly the same as those in TD-DFT \cite{Ullrich11}.

If we apply the Heisenberg equation of motion
\begin{equation}
	\label{eq:heisenberg_motion}
	\frac{\diff O}{\diff t}=\left\langle\frac{\partial \hat{O}}{\partial t}\right\rangle + i\langle[\hatH, \hat{O}]\rangle
\end{equation}
to the Hamiltonians of the KS and the noninteracting boson system, we get \cite{Ullrich11}
\begin{equation}
	\label{eq:Tsvs}
	\frac{\diff T\s(t)}{\diff t}= -\intdbrs\frac{\partial n(\br,t)}{\partial t}v\s(\br,t),
\end{equation}
and (see Appendix \ref{sec:der_Tbvb} for detailed derivation)
\begin{equation}
	\label{eq:Tbvb}
	\frac{\diff T\b(t)}{\diff t}= -\intdbrs\frac{\partial n(\br,t)}{\partial t}v\b(\br,t).
\end{equation}
Subtracting \eqn{eq:Tbvb} from \eqn{eq:Tsvs} we obtain
\begin{equation}
	\label{eq:Tpvp}
	\frac{\diff T\P(t)}{\diff t}= \intdbrs\frac{\partial n(\br,t)}{\partial t}v\P(\br,t).
\end{equation}

\eqn{eq:Tpvp} can be used to check the accuracy of numerical calculations, or as an exact constraint for approximations of $v\P$ as done for other functionals \cite{JMO+20}. This is also the case of its analog in $v\xc$ \cite{LP85} being used in developing XC functionals \cite{KH09}.

\subsubsection{Nonadiabaticity and causality}


It is possible to define time-dependent potentials and kernels from action functionals. Here, we use an analogy of the variational principle method proposed by Vignale \cite{Vignale08} which takes into account the ``causality paradox'' \cite{SD07}.

The action integral of the noninteracting boson system can be written as:
\begin{equation}
	\label{eq:A_b}
	A\b[n]=A_0{}\b[n]-\int_{t_0}^{t_1}\diff t\intdbrs n(\br,t)v\b(\br,t),
\end{equation}
where
\begin{equation}
	\label{}
	A_0{}\b[n]=\int_{t_0}^{t_1}\diff t\,\brakett{\Phi\b[n](t)}{i\frac{\partial}{\partial t}-\hatT}{\Phi\b[n](t)\vphantom{\hat{\frac{\partial}{\partial t}}}}.
\end{equation}
Imposing $A\b$ in \eqn{eq:A_b} to be stationary with respect to variations of the density leads to
\begin{equation}\label{eq:v_b_var}
v\b(\br,t)=\frac{\delta A_0{}\b[n]}{\delta n(\br,t)}-i\braket{\Phi\b[n](t_1)}{\frac{\delta \Phi\b[n](t_1)}{\delta n(\br,t_1)}}.
\end{equation}

Relating the action integrals of the noninteracting boson system and the KS system, we obtain
\begin{equation}
	\label{eq:rel_A_s_A_b}
	A_0{}\b[n]=A_0{}\s[n]+A\P[n],
\end{equation}
and carrying out a similar analysis as \eqn{eq:A_b} through \eqn{eq:v_b_var} for the KS system, plugging in the results of both systems into \eqn{eq:rel_A_s_A_b}, we obtain
\begin{equation}
	\label{funcder}
	v\P(\br,t)=\frac{\delta A\P[n]}{\delta n(\br,t)}+i\braket{\Phi\s[n](t_1)}{\frac{\delta \Phi\s[n](t_1)}{\delta n(\br,t_1)}}-i\braket{\Phi\b[n](t_1)}{\frac{\delta \Phi\b[n](t_1)}{\delta n(\br,t_1)}}.
\end{equation}

The above equation highlights the fact that in order to avoid the so-called ``causality paradox'', the functional derivative of $A\P$ should be augmented by two boundary terms, one stemming from the KS system and one from the noninteracting boson system. This is an analog to the functional derivative of $A\xc$ in conventional TD-DFT augmented by the boundary terms from the interacting system and the KS system. 

\section{Approximating the Pauli kernel}\label{sec:lr}
In this section, we will first derive appropriate Dyson equations relating response functions of the interacting system, the KS system and the noninteracting boson system. In a second step, we analyze the poles of these response functions to better appreciate the nonadiabaticity of the Pauli kernel. Finally, we derive an approximate Pauli kernel and test it in several pilot calculations.

\subsection{Response functions and Dyson equations}


Dyson equations relating  the interacting, KS and the noninteracting boson systems are important as they give us a framework to formulate approximations for the involved kernels \cite{Gould12,NPT+07,NVC09,MZC+04}. 

In linear response theory, we expand the time-dependent density in orders of $v\b$ (where $v\b{}_1$ is the first order potential term) and the first order change in the density, $n_1$, is given by
\begin{equation}\label{}
n_1(\br,t)=\int\diff t'\,\intdbr'\,\chi\b(\br,t,\br',t')v\b{}_1(\br',t'),
\end{equation}
where the density-density response function is
\begin{equation}\label{}
\chi\b(\br,t,\br',t')=\left.\frac{\delta n[v\b](\br,t)}{\delta v\b(\br',t')}\right|_{v\b[n_0](\br)}.
\end{equation}
Defining the time dependent Pauli kernel
\begin{equation}\label{}
f\P(\br,t,\br',t')=\left.\frac{\delta v\P[n](\br,t)}{\delta n(\br',t')}\right|_{n_0},
\end{equation}
and by noticing that both $\chi\b$ and $f\P$ only depend on $t-t'$, they can be represented in frequency space using Fourier transformations 
\begin{equation}\label{}
\chi\b(\br,\br',\omega) = \int \diff(t-t')\,e^{i\omega(t-t')}\chi\b(\br,\br',t-t'),
\end{equation}
and same for $f\P$. We can now introduce the relevant Dyson equations,
\begin{align}
	\label{eq:fp_chi} 
	f\P(\br,\br',\omega)&=\chi\b^{-1}(\br,\br',\omega)-\chi\s^{-1}(\br,\br',\omega),\\
	\label{eq:fphxc}
	f\P{}\Hxc(\br,\br',\omega)&=\chi\b^{-1}(\br,\br',\omega)-\chi^{-1}(\br,\br',\omega),\\
	\label{eq:fhxc}
	f\Hxc(\br,\br',\omega)&=\chi\s^{-1}(\br,\br',\omega)-\chi^{-1}(\br,\br',\omega).
\end{align}

\begin{figure}[tp]
	\centering
	\includegraphics[width=0.70\textwidth]{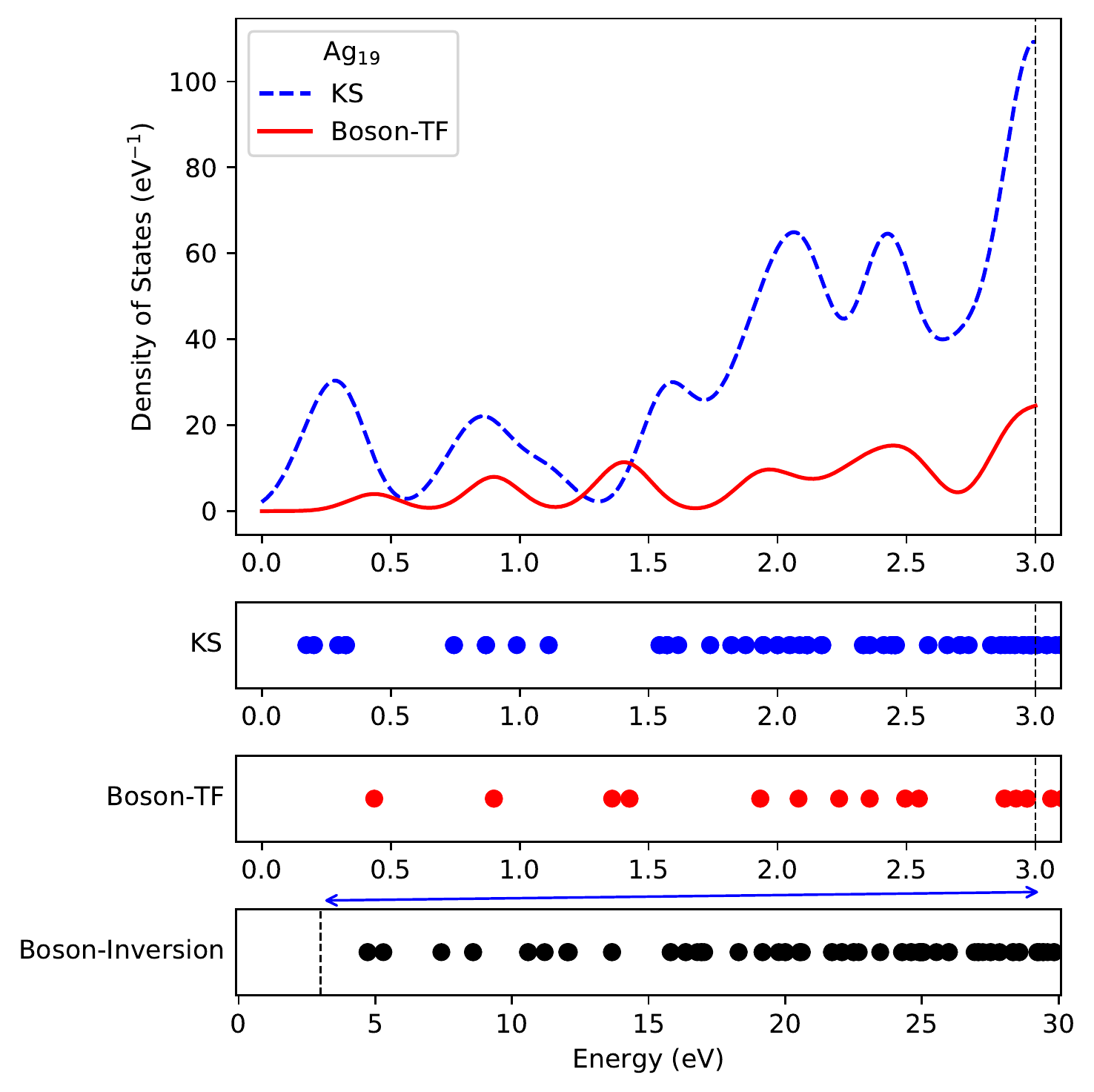}
	\caption{Top panel: Density of states of a Ag$_{19}$ rod (see depiction in inset of Figure \ref{fig:nad_kernel}). Red solid line: noninteracting boson system using the TF potential as $v\P$.  Blue dashed line: KS system. Both systems are computed with the same XC functional (LDA).  Bottom three panels show the energy value of the poles of various response functions. Blue dots: KS system; red dots: approximate noninteracting boson system with TF as $v\P$; black dots: noninteracting boson system with $v\P$ found by inversion of the KS density (see \eqn{eq:invert}). Note: the bottom panel x-axis runs from 0 to 30 eV as opposed to the other panels which run from 0 to 3 eV, as indicated by the black vertical dashed line and the blue arrow.}
	\label{fig:Ag19_poles}
\end{figure}

\eqn{eq:fp_chi} shows the role the Pauli kernel $f\P$ plays in linking the poles of the response functions of the noninteracting boson system and the KS system. Clearly, the response functions of the two systems have poles at different frequency values. To further investigate this, in Figure \ref{fig:Ag19_poles} and Table \ref{tab:poles} we compare the frequency positions of the poles of the noninteracting boson system and the KS system for a Ag$_{19}$ nano-rod \cite{AA18} computed by simply diagonalizing the KS and noninteracting boson Hamiltonians. LDA is used as the XC potential to determine $v\s$, and TF or an exact inversion are used to determine $v\P$. TF is equivalent to employing the Thomas-Fermi-von-Weizs\"acker approximation (TFW) for $T\s[n]$ while the exact inversion of the KS system is carried out {\it via} the following relation
\begin{equation}
\label{eq:invert}
v\b[n](\br)=\frac{\nabla^2 \sqrt{n(\br)}}{2\sqrt{n(\br)}},
\end{equation}
where $n(\br)$ is KS-DFT density. This inversion method is implemented to ensure that the poles of $\chi\b$ and $\chi\s$ derive from the same ground state density.

Figure \ref{fig:Ag19_poles} and Table \ref{tab:poles} shows that for the Ag$_{19}$ system the onset of the poles of the KS response comes at significantly lower frequencies compared to the corresponding poles of the boson response. It is interesting to note that the approximate TF $v\P$ yields excitation energies for the boson system that are much closer to the KS excitations compared to the true boson system computed by inversion.

This demonstrates the role of $f\P$ which is to \textit{red shift} the poles of the boson response towards those of KS response in addition to changing their character by mixing the boson excited states through the Pauli kernel. Unlike the poles of KS response which are generally close to the poles of the interacting response, the differences between poles of boson and KS response seem much larger, highlighting the importance of accounting for nonadiabaticity in $f\P$.

\begin{table}[tp]
	\caption{Comparison of the lowest bosonic and KS pole in eV for the Ag$_{19}$ nano-rod and icosahedral Na$_{55}$ cluster (see Figure \ref{fig:nad_kernel} for a depiction of their structures and Figure \ref{fig:Ag19_poles} for additional information).}
	\label{tab:poles}
	\centering
	\begin{tabular}{cccc}
		\toprule
		System    & KS    & Boson-TF & Boson-inversion \\\midrule
		Ag$_{19}$ & 0.173 &    0.436 & 4.717           \\
		Na$_{55}$ & 0.358 &    0.368 & 3.282           \\\bottomrule
	\end{tabular}
\end{table}


\subsection{Approximating $f\P$ accounting for nonadiabaticity and nonlocality}

The KS response function for the noninteracting uniform electron gas (free electron gas, or FEG, hereafter), aka the frequency-dependent Lindhard function, plays an important role in TD-DFT for deriving approximations to the XC kernel \cite{LGP00,DW00,CDO+98,BLR16,DPG+01,CP07,RNP+20} and is also used as the target response function for parametrizations of nonadiabatic Pauli potentials \cite{WCD+18,NPC+11}. Following a similar paradigm, we first derive the exact response function for the bosonic FEG (BFEG) and then we use the Dyson equation in \eqn{eq:fp_heg} to derive approximations to the Pauli kernel. 

For deriving the BFEG response function, assume $N$ noninteracting bosons confined in a $d$-dimensional cubic cell with the volume of $L^d$, with a uniform external potential. Assuming spin compensation, the BFEG response function can be written as \cite{GV05}:
\begin{equation}\label{eq:chi_nn_heb}
\chi\b(q,\omega)=\frac{1}{L^d}\sum_{\bk}\frac{n_{\bk}-n_{\bk+\bq}}{\omega+\varepsilon_{\bk}-\varepsilon_{\bk+\bq}+i\eta},
\end{equation}
where $n_{\bk}=1/(e^{(\varepsilon_{\bk}-\mu)/T}-1)$ is the Bose-Einstein average occupation of state $\bk$ at temperature $T$ and chemical potential $\mu$. We refer the reader to chapter 4 of Ref.\ \citenum{GV05} for a thorough introduction and derivation of the Lindhard functions in 1, 2 and 3 dimensions. 

\eqn{eq:chi_nn_heb} can be rewritten with a change of variable $\bk\rightarrow\bk-\bq$:
\begin{equation}\label{eq:chi_nn_heb_2}
\chi\b(g,\omega)=\frac{1}{L^d}\left(\sum_{\bk}\frac{n_{\bk}}{\omega+\varepsilon_{\bk}-\varepsilon_{\bk+\bq}+i\eta}+\sum_{\bk}\frac{n_{\bk}}{-\omega+\varepsilon_{\bk}-\varepsilon_{\bk-\bq}-i\eta}\right),
\end{equation}
where the summation is over all occupied states. In the limit of $T\rightarrow0$, only $\bk=0$ states are occupied with the occupation number $N$, therefore we can replace the summation with a multiplication of $N$:
\begin{align}\label{eq:lindhard_b}
\chi\b(q,\omega)&=\frac{N}{L^d}\left(\frac{1}{\omega-q^2/2+i\eta}+\frac{1}{-\omega-q^2/2-i\eta}\right)\nonumber\\
&=\frac{k_F^3}{3\pi^2}\left(\frac{1}{\omega-q^2/2+i\eta}+\frac{1}{-\omega-q^2/2-i\eta}\right),
\end{align}
where the Fermi wavevector, $k_F=(3\pi^2n)^{1/3}$.

\begin{figure}[htp]
	\centering
	\includegraphics[width=\textwidth]{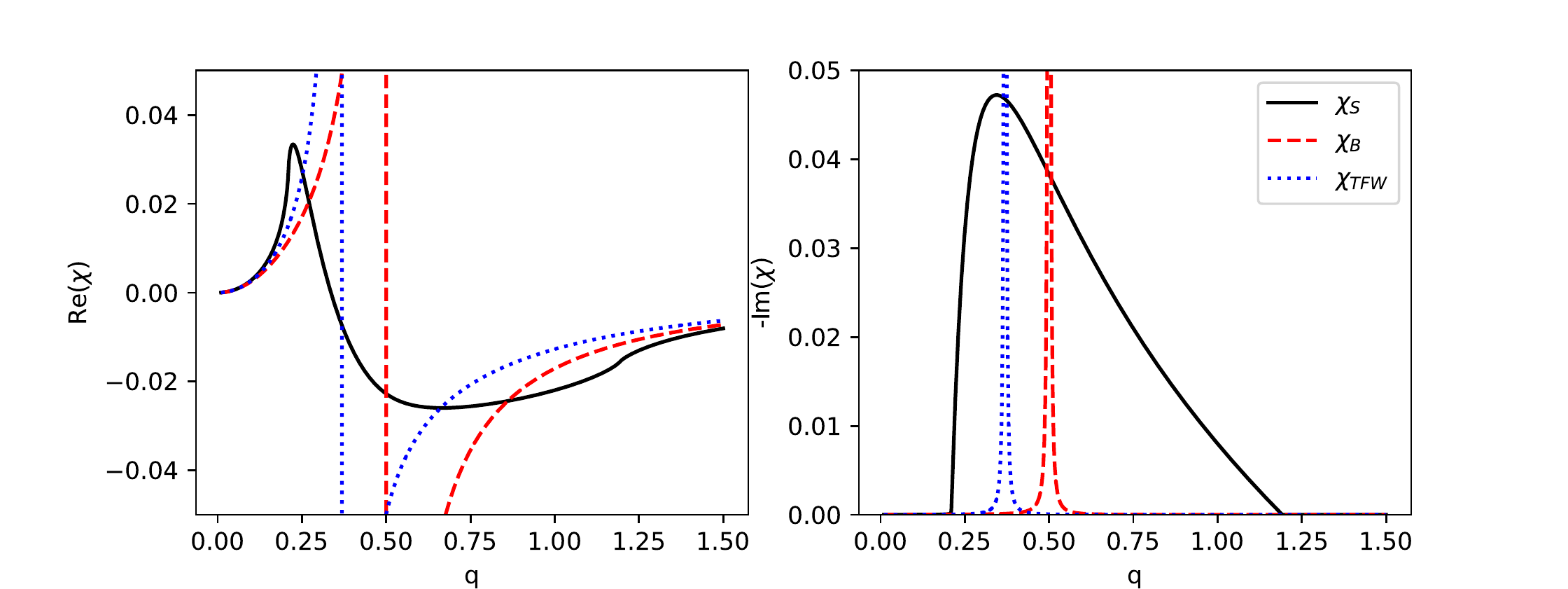}
	\caption{Comparison between the Lindhard function ($\chi\s$ for the FEG), $\chi\b$ for the BFEG (see \eqn{eq:lindhard_b}) and the Thomas-Fermi-von-Weizs\"acker (TFW) response function $\chi{\sss{TFW}}$ (Eq.\ (18) of Ref.\ \citenum{WCD+18}) with respect to $q$ for a uniform density value of $n=0.004$ a.u.\ (corresponding to a Fermi wavevector value of $k_F=0.491$ a.u.), the excitation frequency is set to $\omega=0.125$ a.u.\ and spectral broadening is set to $\eta=0.0001$ a.u..}
	\label{fig:chi0125}
\end{figure}

We note that the formula for $\chi\b$ of the BFEG is simpler than the one for $\chi\s$ of the FEG due to the omission of the integral over $\bk$. Figure \ref{fig:chi0125} shows the comparison between $\chi\s$, $\chi\b$ and the approximate response function for TFW functional $\chi{\sss{TFW}}$ for the FEG. The $\chi{\sss{TFW}}$ response function is given by Eq.\ (18) of Ref.\ \citenum{WCD+18} $\chi{\sss{TFW}}=-(k_F/\pi^2)\left[1+3q^2/4k_F^2-3\omega^2/(k_F^2q^2)\right]^{-1}$ which was also found in Ref.\ \citenum{MBR18}. 
The $\eta\to 0$ limit of the BFEG response in \eqn{eq:lindhard_b} $\lim\limits_{\eta\rightarrow 0}\chi\b=-(k_F^3/3\pi^2)\left[q^2/4-\omega^2/q^2\right]^{-1}$ reproduces the results in Refs.\ \citenum{WCD+18,MBR18} by taking the TF part out of $\chi{\sss{TFW}}$. $\chi\b$ has the same asymptotic behavior as $\chi\s$ for both $q\rightarrow 0$ and $q\rightarrow \infty$. It has a similar shape to $\chi{\sss{TFW}}$ but the singularity is at a different position. Additionally, both $\chi\b$ and $\chi{\sss{TFW}}$ feature vary narrow peak dispersions compared to $\chi\s$. 

Combining \eqn{eq:lindhard_b}, the Lindhard function and the Dyson equation \eqn{eq:fp_chi}, we can generate the exact Pauli kernel for the FEG (in reciprocal space):

\begin{align}\label{eq:fp_heg}
	f\P^{\rm FEG}(q, \omega) =& \frac{3\pi^2}{k_F^3}\left(\frac{1}{\omega-q^2/2+i\eta}+\frac{1}{-\omega-q^2/2-i\eta}\right)^{-1}\nonumber\\
&-\frac{\pi^2q}{k_F^2}\left[\Psi_3\left(\frac{\omega+i\eta}{qk_F}-\frac{q}{2k_F}\right)-\Psi_3\left(\frac{\omega+i\eta}{qk_F}+\frac{q}{2k_F}\right)\right]^{-1},
\end{align}
where
\begin{equation}\label{}
\Psi_3(z)=\frac{z}{2}+\frac{1-z^2}{4}\ln{\frac{z+1}{z-1}}.
\end{equation}

\begin{figure}[htp]
	\centering
	\includegraphics[width=\textwidth]{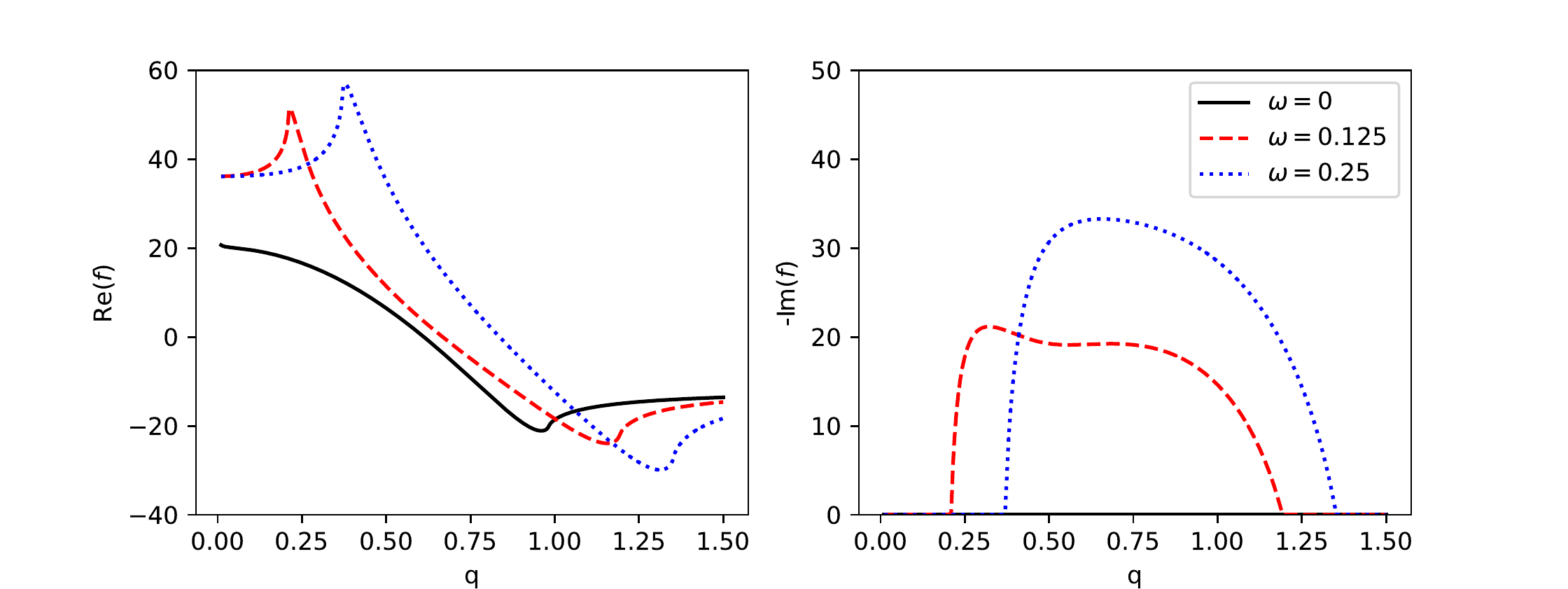}
	\caption{Pauli kernel, $f\P$, for the free electron gas. Plots of $f\P$ with respect to $q$ are shown for different $\omega$ (in a.u.) at $n=0.004$ a.u. ($k_F=0.491$ a.u.) and $\eta=0.0001$ a.u..}
	\label{fig:fp}
\end{figure}

\begin{figure}[htp]
	\centering
	\includegraphics[width=\textwidth]{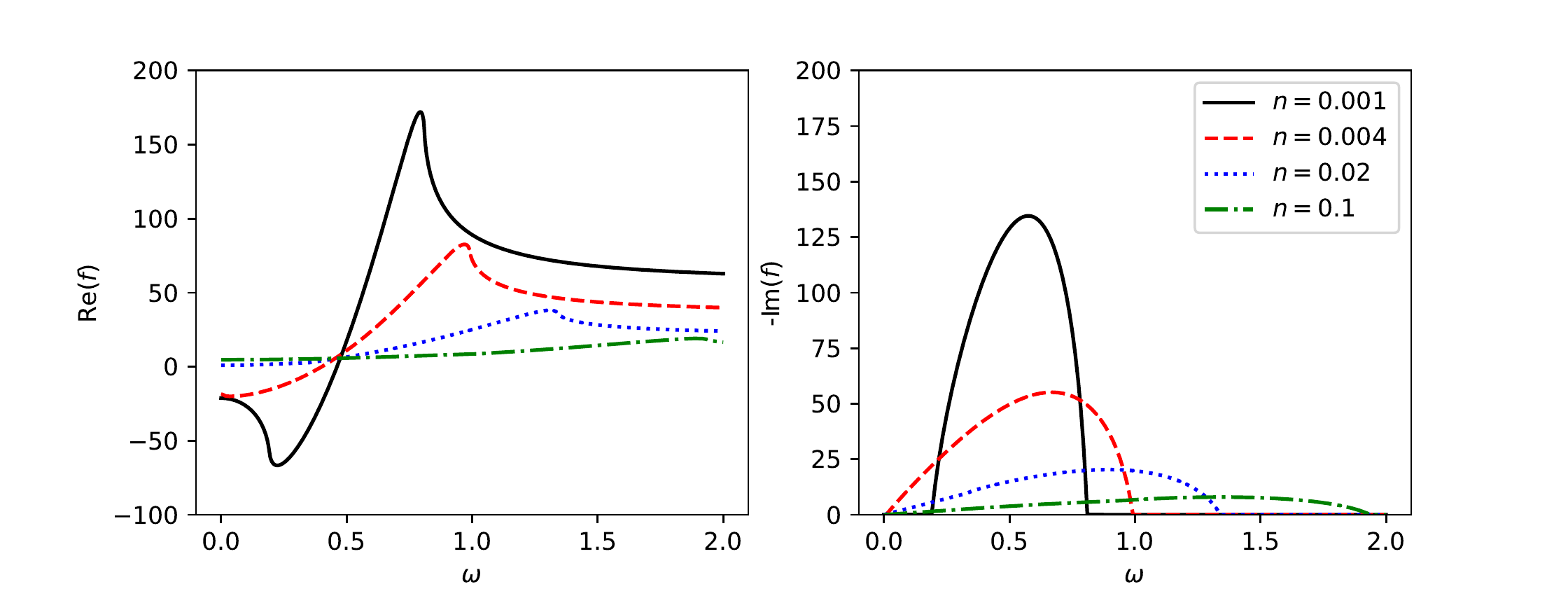}
	\caption{Pauli kernel, $f\P$, for the free electron gas. Plots of $f\P$ with respect to $\omega$ are shown for different $n$ (in a.u.) at $q=1$ a.u. and $\eta=0.0001$ a.u..}
	\label{fig:fp_omega}
\end{figure}


Figure \ref{fig:fp} shows that the Pauli kernel \eqn{eq:fp_heg} follows strongly the features of the Lindhard function.  The real part of $f\P$ has two cusps at the same position of the cusps of the Lindhard function for $\chi\s$, and the imaginary part has a wide feature with the same width as the peak in the Lindhard function for $\chi\s$. Figure \ref{fig:fp_omega} compares the $f\P$-$\omega$ dependence for different choices of density values. $f\P$ gets closer to 0 as the density increases because the system becomes more Thomas-Fermi like \cite{CFL+11}. On the other hand, at lower density, the non-zero features of $f\P$ are more pronounced, for example the feature between $\omega$ from 0.2 a.u. to 0.8 a.u. for $n=0.001$ a.u.. 

\subsection{Extending the Pauli kernel from the uniform electron gas to inhomogeneous systems}\label{sec:nad_kernel}
One way to extend \eqn{eq:fp_heg} to general systems is to replace the Fermi wave vector $k_F$ with a two-body Fermi wave vector $\xi(\br, \br')$, such as
\begin{equation}\label{eq:xi1}
\xi(\br,\br')=\left[\frac{k_F^\gamma(\br)+k_F^\gamma(\br')}{2}\right]^{1/\gamma},
\end{equation}
or a geometric average
\begin{equation}\label{eq:xi2}
\xi(\br,\br')=k_F^{1/2}(\br)k_F^{1/2}(\br').
\end{equation}
\eqn{eq:xi1} is used in functionals of the noninteracting kinetic energy ($T\s[n]$) such as CAT \cite{CAT85} and WGC \cite{WGC99} and has the advantage of $\xi(\br,\br')$ always being positive as long as the density is not $0$ at both $\br$ and $\br'$. However, it leads to a quadratic scaling algorithm because it is not possible to formulate the resulting equations in terms of convolution-like integrals where fast Fourier transform can be applied. With the choice of \eqn{eq:xi2} for the Fermi wavevectors, we propose the non-adiabatic correction to the Pauli kernel in the following form (See Appendix \ref{sec:der_fpnad} for detailed derivation):

\begin{equation}
\label{eq:fp_nad}
f\P^{\rm{nad}}(\omega,q)=\frac{i\pi^3}{12}\left(\frac{6}{\xi^2q}+\frac{q}{\xi^4}\right)\omega+\frac{\pi^2(16-\pi^2)}{4\xi^3q^2}\omega^2+\frac{\pi^2(16-3\pi^2)}{48k_F^5}\omega^2.
\end{equation}
Note the last term in \eqn{eq:fp_nad} is purely local (e.g. does not have a contribution from $q$).

We point out that the procedure discussed here is equivalent to the one presented in Ref.\ \citenum{WCD+18}. We, however, include additional terms in the expansion. As it will be clear below, the additional terms (and particularly terms nonlocal in space) contribute significantly.
The second-order expansion of the kernel in \eqn{eq:fp_nad} features both imaginary and real valued terms. The immediate effects of the real-valued terms is to shift the position of the poles, and the imaginary terms broaden the lineshapes. While in this work we focus mainly on excitation energy shifts, broadening of the peaks also is important to represent the increase of accessible states in the KS system compared to the boson system.

To use the kernel in practical calculations, Casida matrix elements need to be computed, such as,
\begin{equation}
\label{}
	K_{ij}=\intdbr\intdbr'\phi_0(\br')\phi_j^*(\br')f\P^{\rm nad}(\omega,\br,\br')\phi_0^*(\br)\phi_i(\br),
\end{equation}
we need to treat the space dependence of $k_F$ in the following way. For last term in \eqn{eq:fp_nad} which is local, we apply the local-density approximation or LDA. Namely, $k_F\equiv k_F(\br)=(3\pi^2n_0(\br))^{1/3}$. The LDA results in the following commonly employed integrals (e.g., for the so-called ALDA kernel in TD-DFT),
\begin{equation}
\label{}
K_{ij}^{\rm{local}}=\intdbr\intdbr'\phi_0(\br')\phi_j^*(\br')f\P^{\rm{nad,local}}(\omega,\br)\delta(\br'-\br)\phi_0^*(\br)\phi_i(\br).
\end{equation}
For the non-local terms, for example, the second to the last term in \eqn{eq:fp_nad}, we use $k_F=k_F^{1/2}(\br)k_F^{1/2}(\br')$, and
\begin{equation}
\label{}
K_{ij}=\intdbr'\phi_0(\br')\phi_j^*(\br')k_F^{-2/3}(\br')\mathcal{F}^{-1}\left\{\frac{\pi^2(16-\pi^2)}{4q^2}\omega^2\mathcal{F}\{k_F^{-2/3}(\br)\phi_0^*(\br)\phi_i(\br)\}\right\},
\end{equation}
where $\mathcal{F}\{\cdot\}$ and $\mathcal{F}^{-1}\{\cdot\}$ represent forward and inverse Fourier transform, respectively.

\subsection{Pilot calculations involving the nonadiabatic Pauli kernel}
To demonstrate the effect of the nonadiabatic Pauli kernel, we present several pilot calculations involving the solution of the Dyson equation in \eqn{eq:fp_chi} linking the boson and KS response functions. That is, we begin with the boson response function, $\chi\b$, and through \eqn{eq:fp_chi} employing an approximate Pauli kernel we recover an approximate $\chi\s$ which we compare to the exact one.  

The role of the nonadiabatic Pauli kernel is to line up the poles of $\chi\b$ with the poles of $\chi\s$. 
According to the examples provided in Figure \ref{fig:Ag19_poles} and Table \ref{tab:poles}, the Pauli kernel should red shift the bosonic poles by as much as 4.5 eV and 2.8 eV for the silver nano-rod and sodium cluster, respectively. Therefore, we expect $f_p$ and its matrix elements with respect to occupied-virtual orbital products to be large in size and comparable to the orbital energy differences.  

Therefore, it is problematic that the simplest approximation to it [i.e., the TF functional with kernel $f\p^{\rm TF}(\br,\brp)=\frac{10}{9}C_{TF}\delta(\br-\brp)n^{-\frac{1}{3}}(\br)$] is strictly positive. 
From the single pole approximation (i.e., $\omega\s=\sqrt{\omega\b^2+2K\omega\b}$, where $\omega_{\sss{S/B}}$ are the KS/boson orbital excitations and $K$ is the matrix element of the Pauli kernel), we see that the positivity of the kernel leads to a blue shift of the first excited state which is opposite to the sought behavior. For these reasons, we expect the nonadiabatic part of the kernel to play a very significant role.

\begin{figure}[tp!]
	\centering
	\includegraphics[width=\textwidth]{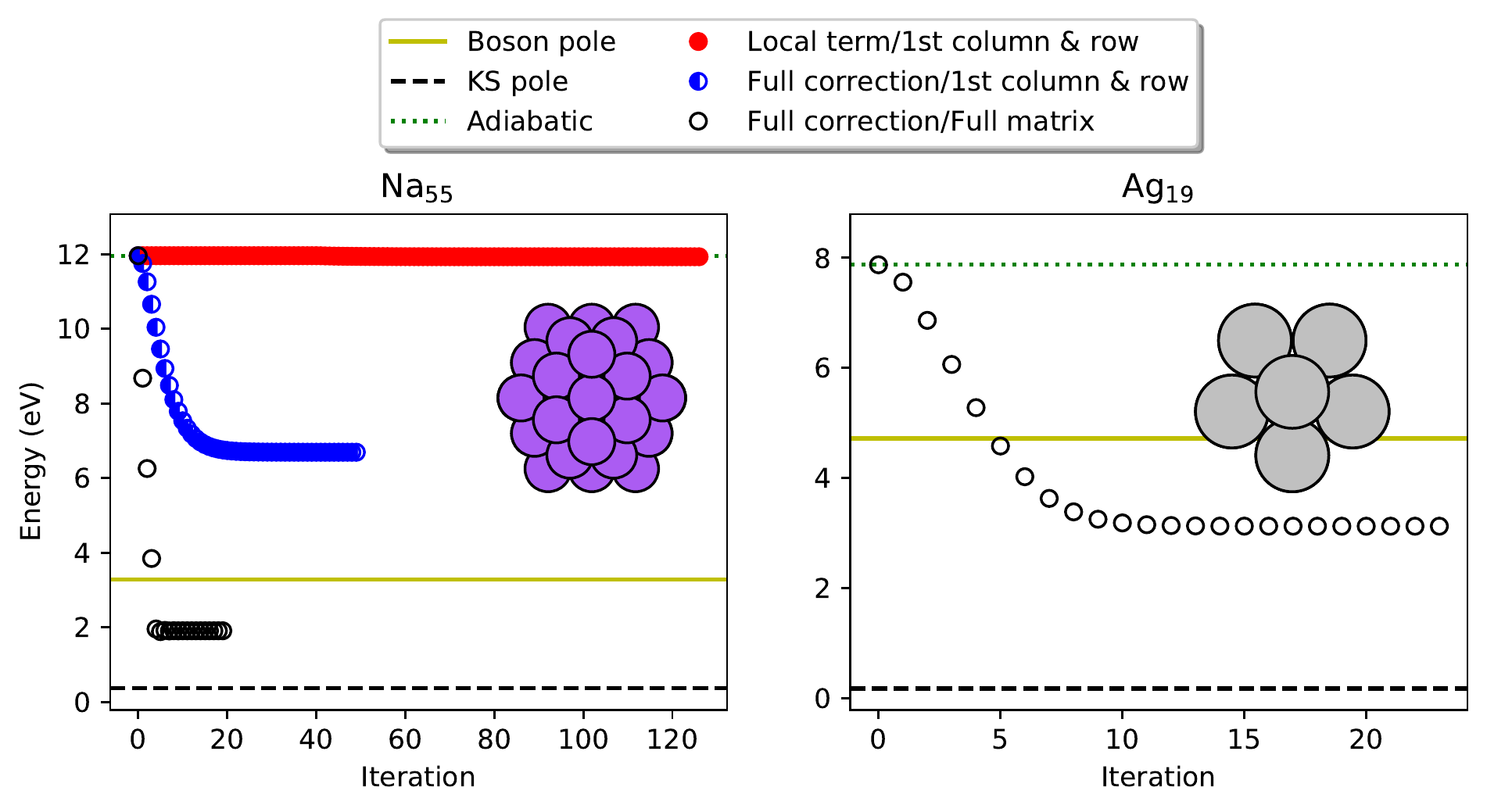}
	\caption{The convergence of the first KS pole with the non-adiabatic $f\P$ correction for an icosahedral Na$_{55}$ cluster and a Ag$_{19}$ nano-rod (structures in inset). The yellow solid line is the position of the bosonic pole, the green dotted line is the position of the KS pole calculated using only the adiabatic $f\P$ and the black dashed line is the position of the KS pole from diagonalizing the KS Hamiltonian. Red filled circles: uses only the local term and couples the first state with the others only. Blue half-filled circles: uses full correction and couples the first state with the others only. Black empty circles: uses full correction and full coupling.}
	\label{fig:nad_kernel}
\end{figure}

In the calculations presented below we start from the bosonic poles calculated by diagonalizing the noninteracting boson Hamiltonian of \eqn{eq:ham_boson}, where the $v\P$ is determined by inversion using \eqn{eq:invert} and therefore it is exact within the numerical precision of the inversion procedure. We then compute the lowest-lying KS pole by solving the Casida equation \cite{Casida95} associated with the Dyson equation \eqn{eq:fp_chi} with an {\it approximate} Pauli kernel and compare to the exact KS pole computed by diagonalization of the KS Hamiltonian. 

We use the adiabatic TF as the adiabatic part of the Pauli kernel and \eqn{eq:fp_nad} evaluated according to the prescriptions given in subsection \ref{sec:nad_kernel} for the nonadiabatic part. The Dyson equation in \eqn{eq:fp_nad} needs to be solved iteratively starting from the adiabatic result. The iterations evolve as follows: (1) at the $n$-th cycle we evaluate the KS response function {\it via} the Dyson equation, $\chi\s^n(\omega_n^{\rm{out}})=\chi\b(\omega_n^{\rm{out}})+\chi\b(\omega_n^{\rm{out}}) f\P(\omega_{n-1}^{\rm{in}})\chi\s^n(\omega_n^{\rm{out}})$; (2) the kernel is updated with the value of the new frequency $\omega_n^{\rm{in}}=\beta \omega_n^{\rm{out}} + (1-\beta)\omega_{n-1}^{\rm{in}}$, and the Dyson equation is solved again to find yet another pole frequency. The $\beta$ mixing parameter was determined adaptively from $\beta=1$ to $\beta=0.1$ to find balance between speed of convergence and stability.


Figure \ref{fig:nad_kernel} shows that the nonadiabatic Pauli kernel derived in subsection \ref{sec:nad_kernel} succeeds in bringing the bosonic pole closer to the KS pole for a sodium cluster, Na$_{55}$ and a silver nanorod, Ag$_{19}$, respectively. For the Na$_{55}$ system, we carried out additional analyses: (1) we show the effect of applying the Pauli kernel on one row/column of the Casida matrix or on the full matrix (blue half-filled circles vs black empty circles) and (2) we show that applying only the space-local part of the nonadiabatic Pauli kernel is not enough to substantially red shift the bosonic pole (red filled circles). 

To ensure that the numerical artifacts in the inverted $v\P$ do not affect the final results, we also performed the same calculations in Figure \ref{fig:nad_kernel} with $v\P$ multiplied by a mask function $m(\br)=1-1/\{1+[n_{\rm{KS}}(\br)/n_{\rm{cutoff}}]^2\}$, where $n_{\rm{KS}}(\br)$ is the KS density used as the input density of the inversion, and $n_{\rm{cutoff}}=10^{-4}$ a.u.. The mask function removed all artifacts from $v\P$. The results of these calculations are almost identical to those in Figure \ref{fig:nad_kernel}.

The above examples show that the proposed nonadiabatic correction to the Pauli kernel provides the required red shift to the bosonic pole towards the KS pole, provided the kernel contains space-nonlocal terms.

\section{Summary}
In the first part of this work, we present orbital-free TD-DFT and derive properties (or conditions) that the Pauli energy, potential and kernel must satisfy. Furthermore, we provide proofs of theorems that also apply to conventional TD-DFT, such as the Runge-Gross and van Leuuwen theorems.

In the second part of this work, we derive an approximation for the Pauli kernel based on the response properties of the uniform electron gas. The derived kernel is nonlocal in space as well as nonlocal in time because it features an explicit frequency dependence. Pilot calculations show that the nonadiabatic part of the kernel is much more important than the adiabatic part and that space-nonlocal terms in the kernel play a fundamental role. The proposed kernel is capable of correctly red shifting the orbital free (bosonic) orbital excitation energies, bringing them close to the KS orbital excitations. While the final result of the excitation energy of the lowest-lying KS excited state for a sodium cluster and a silver nanorod are only in semiquantitative agreement with the exact result, this work sheds light on a path to develop nonadiabatic Pauli kernels for modeling with time-dependent orbital-free DFT systems out of equilibrium in a computationally cheap and semiquantitatively accurate way. 



\appendix
\section{Derivation of \eqn{eq:motion_j}}\label{sec:der_Fkin}
We start with the equation of motion for the current density operator
\begin{align}\label{eq:motion_d}
&i\frac{\partial}{\partial t}\bj(\br,t)\nonumber\\
=&\brakett{\phi\b(t)}{[\hat{\bj}(\br), \hat{H\b}(t)]}{\phi\b(t)}\nonumber\\
=&\brakett{\phi\b(t)}{\hat{\bj}(\br)\hat{H\b}(t)-\hat{H\b}(t)\hat{\bj}(\br)}{\phi\b(t)}\nonumber\\
=&\brakett{\phi\b(t)}{\hat{\bj}(\br)\hat{T}-\hat{T}\hat{\bj}(\br)+\hat{\bj}(\br)\hatV\b(t)-\hatV\b(t)\hat{\bj}(\br)}{\phi\b(t)},
\end{align}

where
\begin{align}\label{eq:motion_dp1}
&\brakett{\phi\b(t)}{\hat{\bj}(\br)\hatV\b(t)-\hatV\b(t)\hat{\bj}(\br)}{\phi\b(t)}\nonumber\\
=&\frac{1}{2i}\sum_l^N\int\diff\br_1\dots\diff\br_N\phi\b^*(\br_1,\br_2,\dots,\br_N,t)\nonumber\\
&\{\nabla_l[\delta(\br-\br_l) v\b(\br_l,t)]+\delta(\br-\br_l)\nabla_l v\b(\br_l,t)-v\b(\br_l,t)\nabla_l\delta(\br-\br_l)-v\b(\br_l,t)\delta(\br-\br_l)\nabla_l\}\nonumber\\
&\phi\b(\br_1,\br_2,\dots,\br_N,t)\nonumber\\
=&\frac{N}{2i}\{\phi\b^*(\br,t)\nabla [v\b(\br,t)\phi\b(\br,t)]-\nabla[\phi\b^*(\br,t)v\b(\br,t)\phi\b(\br,t)]+\phi\b^*(\br,t)\nabla [v\b(\br,t)\phi\b(\br,t)]\nonumber\\
&+\nabla[\phi\b^*(\br,t)v\b(\br,t)\phi\b(\br,t)]-\phi\b^*(\br,t) v\b(\br,t)\nabla\phi\b(\br,t)-\phi\b^*(\br,t) v\b(\br,t)\nabla\phi\b(\br,t)\}\nonumber\\
=&\frac{N}{i}\phi\b^*(\br,t)\phi\b(\br,t)\nabla v\b(\br,t)\nonumber\\
=&\frac{1}{i}n(\br,t)\nabla v\b(\br,t),
\end{align}

and
\begin{align}\label{eq:motion_dp2}
&\brakett{\phi\b(t)}{\hat{\bj}(\br)\hat{T}-\hat{T}\hat{\bj}(\br)}{\phi\b(t)}\nonumber\\
=&-\frac{1}{4i}\sum_l^N\sum_k^N\int\diff\br_1\dots\diff\br_N\phi\b^*(\br_1,\br_2,\dots,\br_N,t)\nonumber\\
&\{\nabla_l[\delta(\br-\br_l) \nabla_k^2]+\delta(\br-\br_l)\nabla_l \nabla_k^2-\nabla_k^2\nabla_l\delta(\br-\br_l)-\nabla_k^2[\delta(\br-\br_l)\nabla_l]\}\nonumber\\
&\phi\b(\br_1,\br_2,\dots,\br_N,t)\nonumber\\
=&-\frac{N}{4i}\{\phi\b^*(\br,t)\nabla^3\phi\b(\br,t)-\nabla[\phi\b^*(\br,t)\nabla^2\phi\b(\br,t)]+\phi\b^*(\br,t)\nabla^3\phi\b(\br,t)\nonumber\\
&-\phi\b^*(\br,t) \nabla^3\phi\b(\br,t)+3\nabla[\phi\b^*(\br,t)\nabla^2\phi\b(\br,t)]-3\nabla^2[\phi\b^*(\br,t)\nabla\phi\b(\br,t)]+\nabla^3[\phi\b^*(\br,t)\phi\b(\br,t)]\nonumber\\
&-\phi\b^*(\br,t) \nabla^3\phi\b(\br,t)+2\nabla[\phi\b^*(\br,t)\nabla^2\phi\b(\br,t)]-\nabla^2[\phi\b^*(\br,t)\nabla\phi\b(\br,t)]\nonumber\\
&+(N-1)\{\phi\b^*(\br,t)\nabla\phi\b(\br,t)-\nabla[\phi\b^*(\br,t)\phi\b(\br,t)]+\phi\b^*(\br,t)\nabla\phi\b(\br,t)\nonumber\\
&+\nabla[\phi\b^*(\br,t)\phi\b(\br,t)]-\phi\b^*(\br,t) \nabla\phi\b(\br,t)-\phi\b^*(\br,t) \nabla\phi\b(\br,t)\}\int\diff\br'\phi\b^*(\br',t)\nabla'^2\phi\b(\br',t)\}\nonumber\\
=&-\frac{N}{4i}\{4\nabla[\phi\b^*(\br,t)\nabla^2\phi\b(\br,t)]-4\nabla^2[\phi\b^*(\br,t)\nabla\phi\b(\br,t)]+\nabla^3[\phi\b^*(\br,t)\phi\b(\br,t)]\}\nonumber\\
=&-\frac{N}{4i}\{-4[\nabla\phi\b^*(\br,t)][\nabla^2\phi\b(\br,t)]-4[\nabla^2\phi\b^*(\br,t)][\nabla\phi\b(\br,t)]+\nabla^3[\phi\b^*(\br,t)\phi\b(\br,t)]\}.
\end{align}

Combining \eqn{eq:motion_d} through \eqn{eq:motion_dp2} we get
\begin{equation}\label{}
\frac{\partial}{\partial t}\bj(\br,t)=-n(\br,t)\nabla v\b(\br,t))-\bF\kin(\br,t),
\end{equation}
where
\begin{equation}\label{}
\bF\kin(\br,t)=\frac{N}{4}\{4[\nabla\phi\b^*(\br,t)][\nabla^2\phi\b(\br,t)]+4[\nabla^2\phi\b^*(\br,t)][\nabla\phi\b(\br,t)]-\nabla^3[\phi\b^*(\br,t)\phi\b(\br,t)]\}.
\end{equation}
\section{Derivation of \eqn{eq:Tbvb}}\label{sec:der_Tbvb}
Plug the Hamiltonian of the noninteracting boson system as $\hat{O}$ into \eqn{eq:heisenberg_motion}, we obtain
\begin{align}
	\label{eq:Tbvb1}
	\frac{\diff E(t)}{\diff t} =& \left\langle\frac{\partial \hatH}{\partial t}\right\rangle\nonumber\\
	=& \left\langle\frac{\partial \hatT}{\partial t}\right\rangle+\left\langle\frac{\partial \hatV\b(t)}{\partial t}\right\rangle\nonumber\\
	=& \intdbrs\frac{\partial v\b(t)}{\partial t}n(t).
\end{align}
From \eqn{eq:e_boson} we obtain
\begin{equation}
	\label{eq:Tbvb2}
	\frac{\diff E(t)}{\diff t}=\frac{\diff T\b(t)}{\diff t}+\intdbrs v\b(\br,t)n(\br,t).
\end{equation}
Combining \eqn{eq:Tbvb1} and \eqn{eq:Tbvb2} we get
\begin{equation}
	\label{eq:Tbvb3}
	\frac{\diff T\b(t)}{\diff t}=-\intdbrs \frac{\partial n(\br,t)}{\partial t}v\b(\br,t).
\end{equation}
\section{Derivation of non-adiabatic correction to the Pauli kernel}\label{sec:der_fpnad}
First we replace variables $\omega$, $q$, $\eta$ with their dimensionless counterparts $\bar{\omega} = \omega/qk_F$, $\bar{\eta}=q/2k_F$, $\bar{\gamma}=\eta/qk_F$. We can rewrite $\chi\b$ and $\chi\s$
\begin{equation}
	\label{}
	\chi\b(\bar{\omega},\bar{\eta},\bar{\gamma})=\frac{6\pi^2\bar{\eta}}{k_F}\left(\frac{1}{\bar{\omega}-\bar{\eta}+i\bar{\gamma}}+\frac{1}{-\bar{\omega}-\bar{\eta}-i\bar{\gamma}}\right)^{-1},
\end{equation}
\begin{equation}
	\label{}
	\chi\s(\bar{\omega},\bar{\eta},\bar{\gamma})=\frac{2\pi^2\bar{\eta}}{k_F}\left[\Psi_3(\bar{\omega}-\bar{\eta}+i\bar{\gamma})-\Psi_3(\bar{\omega}+\bar{\eta}+i\bar{\gamma})\right]^{-1}.
\end{equation}
Using the Dyson equation \eqn{eq:fp_chi} that relates $\chi\b$ and $\chi\s$, we find the non-adiabatic and adiabatic Pauli kernel $f\P$ and $f\P{}_0$ as following
\begin{equation}
	\label{}
	f\P(\bar{\omega},\bar{\eta},\bar{\gamma})=\chi\b(\bar{\omega},\bar{\eta},\bar{\gamma})-\chi\s(\bar{\omega},\bar{\eta},\bar{\gamma}),
\end{equation}
\begin{equation}
	\label{}
	f\P{}_0(\bar{\eta})=\chi\b(\bar{\omega}=0,\bar{\eta},\bar{\gamma}=0)-\chi\s(\bar{\omega}=0,\bar{\eta},\bar{\gamma}=0).
\end{equation}
Expand $f\P$ and $f\P{}_0$ around $\bar{\eta}=0$ up to the second order of $\bar{\eta}$, we obtain
\begin{align}
	\label{}
	f\P(\bar{\omega},\bar{\eta},\bar{\gamma})=&\frac{\pi^2}{k_F}\bigg\{\left[-3\bar{\gamma}^2+6i\bar{\gamma}\bar{\omega}+3\bar{\omega}^2
	+\frac{2i}{2i+(\bar{\gamma}-i\bar{\omega})\log(\frac{1+i\bar{\gamma}+\bar{\omega}}{-1+i\bar{\gamma}+\bar{\omega}})}\right]\nonumber\\
	&+\left[-3-\frac{4}{3[\bar{\gamma}-i(-1+\bar{\omega})]^2[\bar{\gamma}-i(1+\bar{\omega})]^2[2i+(\bar{\gamma}-i\bar{\omega})
		\log(\frac{1+i\bar{\gamma}+\bar{\omega}}{-1+i\bar{\gamma}+\bar{\omega}})]^2}\right]\bar{\eta}^2\nonumber\\
	&+O(\bar{\eta}^4)\bigg\},
\end{align}
\begin{equation}
	\label{}
	f\P{}_0(\bar{\eta})=\frac{\pi^2}{k_F}\left[1-\frac{8}{3}\bar{\eta}^2+O(\bar{\eta}^3)\right].
\end{equation}
Take the limit $\bar{\gamma}\rightarrow 0$ and expand around $\bar{\omega}=0$ up to the second order of $\bar{\omega}$, we obtain
\begin{equation}
	\label{}
	f\P(\bar{\omega},\bar{\eta})=\frac{\pi^2}{k_F}\left[\left(1-\frac{8}{3}\bar{\eta}^2\right)+\frac{i\pi}{6}(3+2\bar{\eta}^2)\bar{\omega}+\frac{1}{12}(48-3\pi^2+16\bar{\eta}^2-3\pi^2\bar{\eta}^2)\bar{\omega}^2+O(\bar{\omega}^3)\right].
\end{equation}
The non-adiabatic correction to the Pauli kernel is
\begin{align}
\label{}
f\P^{\rm{nad}}(\bar{\omega},\bar{\eta})=&f\P(\bar{\omega},\bar{\eta})-f\P{}_0(\bar{\eta})\nonumber\\
=&\frac{\pi^2}{k_F}\bigg[\frac{i\pi}{6}(3+2\bar{\eta}^2)\bar{\omega}+\frac{1}{12}(48-3\pi^2+16\bar{\eta}^2-3\pi^2\bar{\eta}^2)\bar{\omega}^2\bigg].
\end{align}

We note that exchanging the order of the $\bar \omega$ and $\bar \eta$ limits does not affect the final form of the kernel because the $\bar\omega\to 0$ limit must occur for a finite $q$. This is in contrast to taking the limit with respect to $\omega$ and $q$ where the non-analiticity of the Lindhard function in terms of these variables results in a limit dependent on the order of operations. \cite{GV05}

Replace the variables back, we get
\begin{equation}
\label{eq:fpnad_appendix}
f\P^{\rm{nad}}(\omega,q)=\frac{i\pi^3}{12}\bigg(\frac{6}{k_F^2q}+\frac{q}{k_F^4}\bigg)\omega+\frac{\pi^2(16-\pi^2)}{4k_F^3q^2}\omega^2+\frac{\pi^2(16-3\pi^2)}{48k_F^5}\omega^2.
\end{equation}

\begin{acknowledgments} 
This work is supported by the U.S.\ Department of Energy, Office of Basic Energy Sciences, under Award Number DE-SC0018343. The authors acknowledge the Office of Advanced Research Computing (OARC) at Rutgers, The State University of New Jersey for providing access to the Amarel and Caliburn clusters and associated research computing resources that have contributed to the results reported here. URL: http://oarc.rutgers.edu
\end{acknowledgments}
\bibliography{td-of-dft}
\end{document}